# Structure of Germanene/Al(111): a Two-Layers Surface Alloy

*K. Zhang,[#] D. Sciacca,[†] M.-C. Hanf,[&§] R. Bernard,[#] Y. Borensztein,[#] A. Resta,[♦] Y. Garreau,[♦‡] A. Vlad,[♦] A. Coati,[♦] I. Lefebvre,[†] M. Derivaz,[&§] C. Pirri,[&§] P. Sonnet,[&§] R. Stephan,[&§] G. Prévot[#*]*

[#] Sorbonne Université, Centre National de la Recherche Scientifique, Institut des NanoSciences de Paris, INSP, F-75005 Paris, France

[†] Univ. Lille, CNRS, Centrale Lille, Univ. Polytechnique Hauts-de-France, Junia-ISEN, UMR 8520 - IEMN, F-59000 Lille, France

[&] Université de Haute Alsace, CNRS, IS2M UMR7361, F-68100 Mulhouse, France

[§] Université de Strasbourg, France

[♦] Synchrotron SOLEIL, L'Orme des Merisiers Saint-Aubin, BP 48 91192 Gif-sur-Yvette Cedex, France

[‡] Université de Paris, Laboratoire Matériaux et Phénomènes Quantiques, CNRS, F-75013, Paris, France

ABSTRACT

Unlike silicene, for which the demonstration of the existence has been done through numerous independent studies, the possibility of growing epitaxial germanene remains highly controversial. It has been recently shown by scanning tunneling microscopy that the $(3 \times 3)$ surface reconstruction formed upon Ge deposition on Al(111) presents a honeycomb structure, and it was assigned to a pure germanene monolayer. Using quantitative measurements by surface X-ray diffraction compared to density functional theory calculations, we demonstrate that this Ge/Al(111) $(3 \times 3)$ reconstruction corresponds in fact to a mixed Ge-Al honeycomb layer on top of an alloyed interfacial layer. The model of a germanene monolayer on top of the Al(111) surface can be completely excluded.

## Introduction

Since the discovery of the peculiar electronic properties of graphene, an important effort has been undertaken to synthesize other 2D materials.[1] Among them, germanene, a 2D honeycomb sheet of germanium atoms, attracts a great interest due to its electronic structure similar to the one of graphene. As compared to graphene, the predicted germanene layer presents a low-buckled structure and a mixed sp2-sp3 hybridization.[2] The Dirac cone is however preserved at the K point of the Brillouin zone.[2] The high spin-orbit coupling for Ge atoms leads to a gap opening at the Fermi level, of the order of 24 meV at the Dirac point,[3] that is much larger than in graphene (< 0.05 meV). Higher quantum spin Hall effects and the possibility of energy band gap engineering are thus expected for germanene, opening interesting perspective for valley- and spin-dependent quantum Hall conductivity and integration of germanene in electronic devices.

As germanene cannot be obtained by exfoliation from a bulk material, it has to be grown on a substrate. Germanene synthesis was first reported in 2014 on Pt(111),[4] and then on GePt(110),[5] Au(111),[6] Al(111)[7], HOPG,[8] AlN,[9] $MoS_2$,[10] or Cu(111).[11] In spite of the high number of studies performed, the existence of germanene has remained highly controversial,[12] in particular as compared with silicene where there is now a consensus on the structures that form on Ag(111). These silicene layers correspond to low-buckled honeycomb structures with a negligible in-plane lattice distorsion.[13–15] On the contrary, as compared with the lattice constant (0.397 nm – 0.402 nm) for free-standing germanene,[2,16] much larger distortions and misfits have been experimentally found for epitaxial germanene. A large number of structures are hypothesized with a lattice constant in the 0.43 – 0.44 nm range, corresponding thus to a high misfit of 7-10%. This is the case for germanene/Pt(110),[5] germanene/Au(111),[6] $(2 \times 2)$ germanene/ $(3 \times 3)$Al(111),[7,17–21] and germanene/HOPG.[8] On the contrary, much smaller lattice constants, in the order of 0.38 nm, have been obtained in a framework of honeycomb structures for germanene/$MoS_2$[10] and for a $(2 \times 2)$ germanene/$(\sqrt{7} \times \sqrt{7})$R19.1°Al(111).[19] However, for this latter structure, a model of a $(\sqrt{3} \times \sqrt{3})$ germanene reconstruction[22] has also been proposed, corresponding to a lattice constant of 0.437 nm.

Among the various substrate mentioned above for the germanene growth, Al(111) has been the most studied system up to now. If a large number of experiments have been interpreted in the framework of a honeycomb germanene layer, the possible formation of alloyed phases has been also raised,[23–25] similar to what is observed for Ge growth on Au(111)[26,27] or on Ag(111).[28,29] In that case, the different proposed models correspond either to a honeycomb germanene layer on an $Al_2Ge$ interfacial layer,[23] or to a honeycomb layer with 3 Al atoms and 5 Ge atoms,[24] or two layers thick alloys.[25]

In order to solve the structure of germanene/Al(111), scanning tunneling microscopy (STM) experiments have been performed very recently with an unprecedented resolution.[21] All surface atoms can be identified on the images, confirming models of 8 atoms arranged in a $(2 \times 2)$ honeycomb for $(3 \times 3)$Al(111) and 6 atoms arranged in a $(\sqrt{3} \times \sqrt{3})$ honeycomb for $(\sqrt{7} \times \sqrt{7})$R19.1° Al(111). From a careful comparison with density functional theory (DFT) simulations, it was concluded that these surface atoms are most probably Ge atoms.

In this paper, using surface X-ray diffraction (SXRD) compared with DFT calculations, we demonstrate that the germanene/Al(111) $(3 \times 3)$ reconstruction corresponds in fact to a mixed honeycomb layer on top of an alloyed interfacial layer.

## Methods

SXRD experiments were performed at the SIXS beamline of SOLEIL synchrotron. The Al(111) sample was prepared by repeated cycles of Ar+ sputtering and annealing at T=750 K. Ge was evaporated in the diffraction chamber from a crucible using a commercial Omicron Nanotechnology e-beam evaporator with a sample kept at ~ 373 K. The Ge flux was kept constant during evaporation with a deposition rate of ~ 1.4 ML/h. The sample was analyzed with 18.46 keV X-rays at a grazing incidence angle of 0.2°. Diffracted X-rays were detected by an hybrid pixel detector.[30] The diffracted intensity was measured by performing a combination of rocking scans and continuous "L-scans" along the crystal truncation rods. A reference scan was taken periodically to correct the intensity from the variations due to the layer degradation, which was fitted with a linear decay with time constant $\tau = 0.02\text{h}$. We used the "binoculars" software to produce three-dimensional (3D) intensity data in the reciprocal space from the raw data.[31] The intensity was further integrated along the direction parallel to the surface to get the structure factors. For this purpose, the data were fitted with the product of a lorentzian lineshape in one direction with a gaussian lineshape convolved with a door lineshape in the other direction, using a home-made software. We finally obtained a set of 1274 structure factors along 19 inequivalent reconstruction rods.

DFT calculations have been performed by means of the Vienna Ab Initio Simulation Package (VASP) code[32–35] with the Projector Augmented plane-Wave (PAW) method.[36,37] The electron-electron exchange correlation interactions are described within the Generalized Gradient Approximation (GGA), using the Perdew, Burke, and Ernzerhof (PBE) functional.[38,39] The cutoff energy value is 450 eV, and relaxation is stopped when the components of the forces are lower than 0.05 eV.nm$^{-1}$. The Brillouin zone is sampled with (5 × 5 × 1) k-points. Taking into account the van der Waals interactions did not modify the relative stability of the different models. The unit cell corresponds to a $(3 \times 3)$ Al(111) surface mesh, with a lattice parameter of 0.8570 nm. The slab consists of a surface layer, an interface layer and 9 planes of 9 Al atoms to mimic the bulk part of the Al(111) sample. The surface layer is either a substitutional alloy or contains 8 atoms (Ge atoms only or Ge and Al atoms) arranged in a honeycomb lattice. The interface layer is either a pure Al plane or a substitutional alloy. All atoms were allowed to relax during calculations, except those belonging to the bottom Al plane, whose position was kept fixed. The atomic structures are presented using the visual dynamic software developed by the Theoretical and Computational Biophysics Group in the Beckman Institute for Advanced Science and Technology at the University of Illinois at Urbana Champaign[40,41].

## Experimental results

SXRD is a powerful and quantitative probe of the structure of crystalline surfaces and epitaxial 2D layers. The diffracted intensity from a 2D crystalline layer shows, as function of the momentum transfer, a sharp scattering line-shape parallel to the surface at integer $(H, K)$ indexes of the surface mesh, and a continuous distribution in the out-of-plane direction $(L)$, along so-called reconstruction rods. The layer structure can then be solved by a fine analysis of the intensity measured along these rods.[42]

In the following, the $(H, K, L)$ indices used for indexing a reflection in reciprocal space refer to the Al(111) $(3 \times 3)$ reconstruction basis ($a = 0.8590$ nm, $b = 0.8590$ nm, $c = 0.7014$ nm, $\alpha = \beta = 90°, \gamma = 120°$). During Ge deposition on the Al substrate held at 373 K, we have followed by SXRD the evolution of the intensity for the (2, 0, 0.1) reflection. As soon as evaporation starts, a diffraction peak appears, indicating that $(3 \times 3)$ reconstructed domains have formed on the surface. The evaporation was stopped at intensity saturation of this peak. After evaporation, the full width at half maximum (FWHM) of the peaks at $(2, 0, 0.1)$ and $(0, 2, 0.1)$ is $\Delta q = 0.045$ nm$^{-1}$. This corresponds to a mean domain size of 140 nm.

In Fig. 1 are presented the experimental structure factors $(F_{\mathrm{exp}})$ measured along the $(10L)$ and $(11L)$ rods. As can be seen, along the different rods, $F_{\mathrm{exp}}(L)$ display periodic oscillations with a period equal to 2. They can be schematically associated with interfering objects at a distance of 0.35 nm along the direction normal to the surface. This cannot correspond to a model of a germanene layer on a flat Al(111) substrate. Indeed, for a free-standing germanene layer with a buckling equal to 0.064 nm, one would expect oscillations with a period equal to 11.

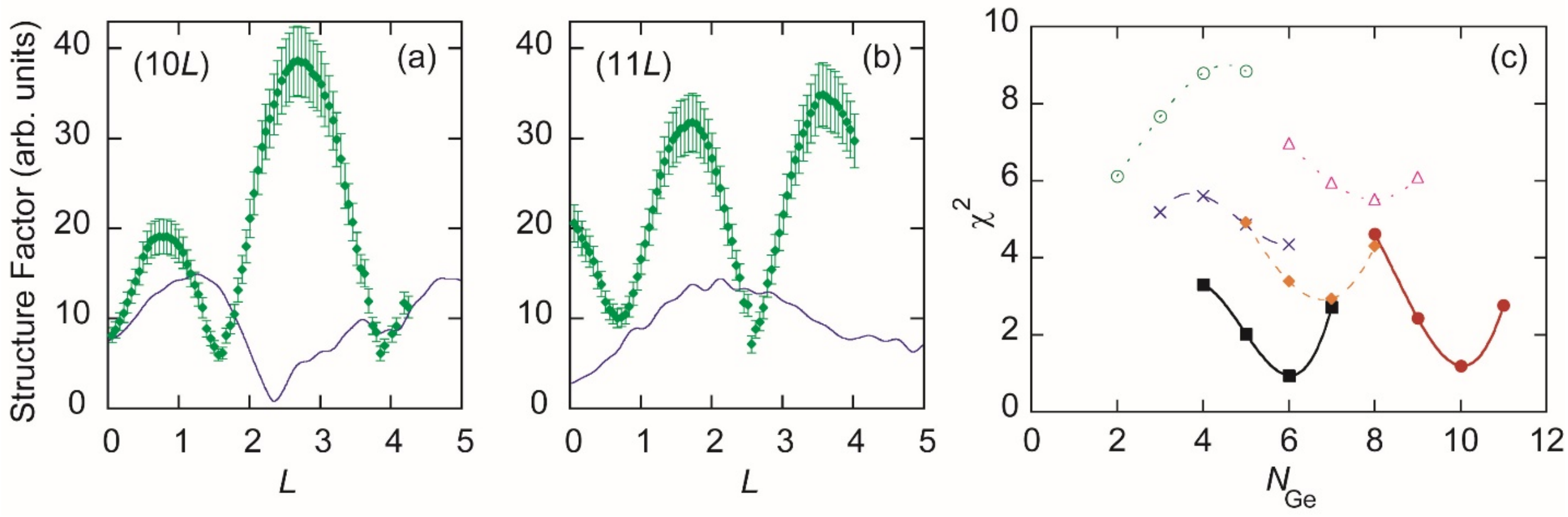

**Figure 1.** (a,b) Structure factors measured along the (10L) and (11L) rods. Comparison between experiments (green lozenges) and simulations (blue line) with a model of 1H germanene. (c) Evolution of the $\chi^2$ with $N_{\mathrm{Ge}}$, the total number of Ge atoms per unit cell for different models. Black squares: $Ge_4Al_4/Ge_nAl_{9-n}$; red dots: $Ge_8/Ge_nAl_{9-n}$; orange lozenges: $Ge_5Al_3/Ge_nAl_{9-n}$; blue crosses: $Ge_3Al_5/Ge_nAl_{9-n}$; green circles: $Ge_2Al_6/Ge_nAl_{9-n}$; pink triangles: $Ge_6Al_2/Ge_nAl_{9-n}$. n is the number of Ge atoms in the interfacial layer.

This indicates that the diffraction signal comes from at least two layers. This could be, for example, due to a large relaxation of the Al plane below the germanene layer, and/or to a much higher buckling of the germanene layer. We have first compared the experimental structure factors with the model of an epitaxial germanene layer, with one Ge sitting at a higher *z* value, corresponding to the 1H model presented in ref. [43] (see Fig. S2m). For this purpose, we have used the atomic positions given by the DFT calculations and a limited set of free parameters: a scale factor and Debye-Waller (DW) factors along the directions perpendicular and planar to the surface.

DW factors describe the structure factors attenuation due to atomic vibrations and varies as $e^{-\frac{B_{//}Q_{//}^2+B_{\perp}Q_{\perp}^2}{16\pi^2}}$ where $Q_{//}$ and $Q_{\perp}$ are the components of the transferred wave-vectors in the directions parallel and perpendicular to the surface, and $B_{//}$ and $B_{\perp}$ are the corresponding DW parameters. These DW parameters were set independent for the different groups of symmetrical atoms of the first two planes, and were the same for all other atoms in the bulk. Their values were restricted to less than 0.1 $nm^2$. The agreement between experimental and simulated ($F_{\text{th}}$) structure factors is estimated by the value of $\chi^2 = \frac{1}{N_{\text{pts}}-N_{\text{par}}}\sum_{N_{\text{Pts}}}\left(\frac{F_{\text{th}}-F_{\text{exp}}}{\sigma_{\text{exp}}}\right)^2$ where $N_{\text{pts}} = 1274$ is the number of experimental structure factors, $N_{\text{par}}$ is the number of free parameters and $\sigma_{\text{exp}}$ is the experimental uncertainty, which takes into account the statistical uncertainty given by the number of counted photons and an overall 10% uncertainty. There is no agreement (see Fig. 1) with the germanene-1H model, with $\chi^2 = 26.6$, showing that even by taking into account the relaxations within the Al substrate, the features observed on the diffraction rods cannot be reproduced by this model.

Consequently, in order to determine the structure of the surface, we have tested all possible configurations involving as well Al and Ge atoms in a bilayer configuration on top of a Al(111) substrate. For the first layer (surface plane), we have explored any position of the atoms that respects the space group symmetries, assuming at least 3 Ge atoms and at most 9 atoms in the surface layer. We have assumed that the space group was the same as the one of the substrate, *i.e.* p3m1. For the second layer (interfacial plane), we have started from an Al(111) plane where zero, one, two or three Al atoms were replaced with Ge atoms, also respecting the p3m1 symmetries (see Fig. S1a-c).[44] We have thus obtained a set of 1888 configurations. For each configuration, the best fit of the structure factors has been obtained by exploring the space of free parameters (an overall scale factor, atomic positions of the first three layers and DW parameters) using the genetic algorithm implemented in SciPy.[45] The agreement was obtained by minimizing $(N_{\text{pts}} - N_{\text{par}})\chi^2 + E$ , where $E$ is a dimension-less Lennard-Jones interaction energy between nearest-neighbor atoms (see Supporting Information for the detail of the procedure). This ensures that unphysical configurations, for example with too short neighbor distances, are excluded. After relaxation, two configurations were found to give an almost perfect agreement with the simulated structure factors. The slightly better fit corresponds to a $Ge_4Al_4$ layer on top of a $Ge_2Al_7$ plane ($\chi^2 \approx 0.9$), whereas the other model corresponds to a $Ge_8$ layer, on top of a $Ge_2Al_7$ plane ($\chi^2 \approx 1.2$) (see Fig. S1c-e). They both display the same structure, with atoms being either Al or Ge atoms. In this structure, 8 atoms in the surface form a honeycomb lattice, as for the germanene-1H model, and 2 Ge atoms replace Al atoms of the second plane. In Fig. 1c is drawn the evolution of $\chi^2$ with $N_{Ge}$, the total number of Ge atoms in the relaxed structure, for different surface compositions of a honeycomb layer and the interfacial layer. The minima obtained for the $Ge_4Al_4/Ge_2Al_7$ ($N_{\text{Ge}}$=6) and for the $Ge_8/Ge_2Al_7$ ($N_{\text{Ge}}$=10) models are clearly visible.

Thus, SXRD results demonstrate that the surface layer is a $(3 \times 3)$ honeycomb layer with 8 atoms per unit cell. This is in very good agreement with the recent high resolution STM observations of Muzychenko et al.[21] However, contrary to their conclusions, our analysis shows that this honeycomb layer sits on top of an alloyed plane, as initially proposed by Fang et al.[23] Moreover, in all models giving a good agreement, the highest atom is a Ge atom above an Al atom, which resides also at a higher $z$ value as compared to the other atoms of the second layer, whereas Muzychenko et al. concluded that the uplifted Ge atom was on a threefold fcc position.[21] It must be underlined that information on atomic positions below the surface is not easily obtained by STM, whereas SXRD gives information on all atoms near the surface. However, in this particular case, two different models with close structure give nearly the same agreement with the experiments, and therefore, SXRD alone cannot answer whether the honeycomb layer is pure germanene or a Ge-Al alloy. Moreover, as SXRD gives only structural information and no energetic information, some of the models obtained from the previous analysis may have a very high surface energy. The so-obtained atomic positions have therefore to be relaxed by DFT calculations in order to minimize the surface energy for each model. After relaxation, the final relaxed atomic positions can then be in poorer agreement with the SXRD data. This is for example the case of the $Ge_8/Al_9$ model (also called 1H germanene): the value of $\chi^2$ obtained by letting atomic positions free to move is 4.6 (see Fig. 1c), whereas the value of $\chi^2$ obtained from the same model relaxed by DFT is 26.6. Consequently, in order to determine the nature of the layer, the various models corresponding to reasonable fits of the SXRD data, as determined from Fig. 1, were further relaxed by DFT.

**DFT computations**

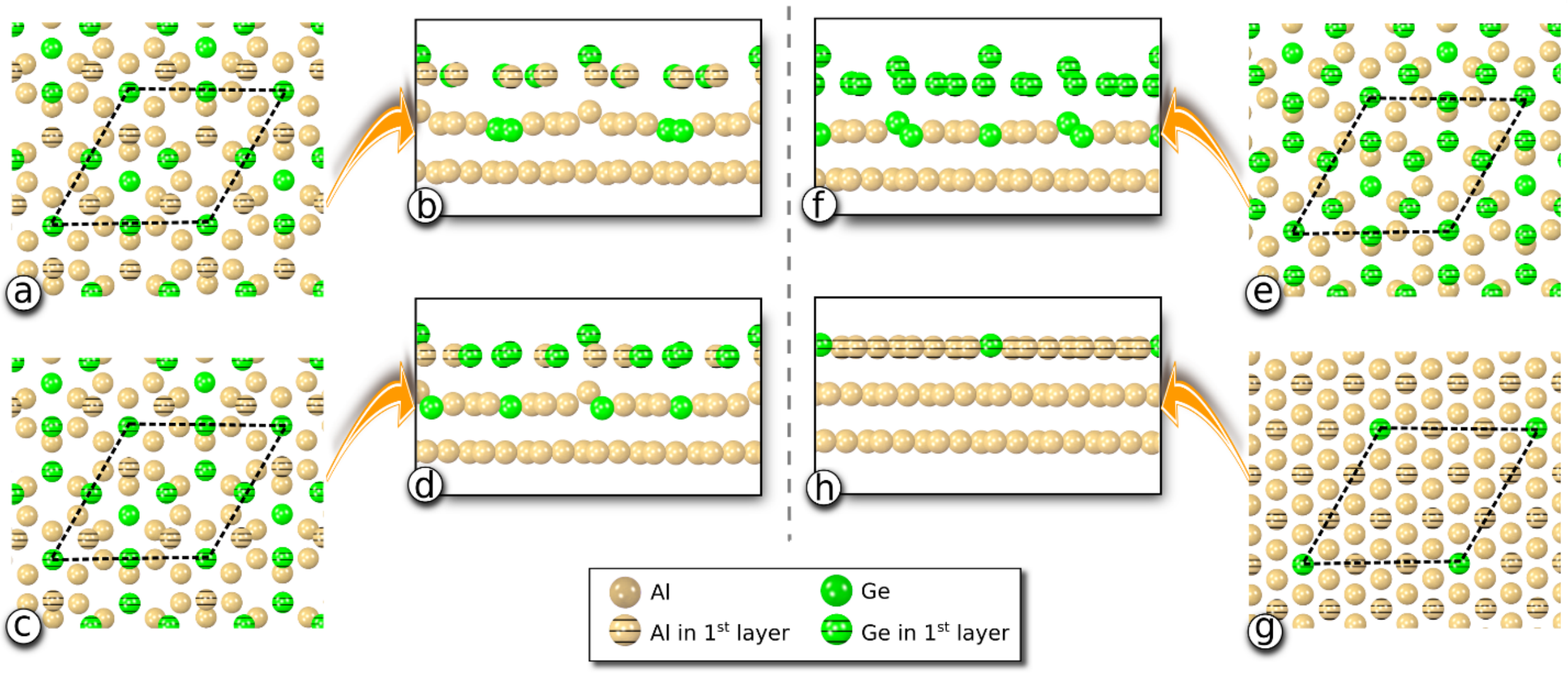

**Figure 2.** Models of Ge layers on Al(111) relaxed by DFT. (a-b) $Ge_4Al_4/Ge_2Al_7$, (c-d) $Ge_5Al_3/Ge_2Al_7$, (e-f) $Ge_8/Ge_3Al_6$, (g-h) $Ge_1Al_8/Al_9$. The side views (b, d, f, h) correspond to a projection along a plane 7° off from the $(11\bar{2})$ plane.

The $Ge_pAl_{8-p}/Ge_nAl_{9-n}$ structures giving a good agreement with SXRD structure factors, as shown in the previous paragraph, were thus relaxed by DFT. Fig. 2a-f presents the $Ge_4Al_4/Ge_2Al_7$, $Ge_5Al_3/Ge_2Al_7$ and $Ge_8/Ge_3Al_6$ configurations obtained after relaxation. The whole set of relaxed configurations is shown in Fig. S2, e-k and m-p. It appears that these systems are very similar. Indeed, the 8 atoms in the surface plane are arranged in a honeycomb lattice, with one uplifted atom with respect to the others. The buckling is in the 0.092-0.131 nm range, and the highest atom (a Ge atom for all models) is located above an Al atom that is also slightly uplifted (about 0.065 nm, except for the $Ge_8/Ge_2Al_7$ model with a value of 0.015 nm). The other atoms in the surface mesh, which are either Ge or Al atoms, are located at about 0.243-0.250 nm above the below-lying plane.

In order to investigate the relative stability of the different systems, we performed a more extensive thermodynamic study for various models, including the best configurations obtained from the SXRD measurements. Three families were studied, namely a pure germanene layer on a mixed Ge-Al surface, a germanene layer with several Ge atoms that are substituted by Al atoms, deposited on a mixed Ge-Al surface and finally a surface alloy made of an hexagonal Al(111) surface plane with several Al atoms substituted with Ge (an example is given in Fig. 2g-h). The latter is located above a pure Al or Ge-Al mixed plane. Since the systems differ not only by their structure but also their chemical composition, it is not easy to simply compare their total energies. We have computed two quantities: the adsorption energy per Ge atom with respect to the bulk Ge energy $E_{\mathrm{Ge\,bulk}}$:

$$E_{ad} = (E_{\mathrm{Ge-Al}} - E_{\mathrm{Al}} - N_{\mathrm{Ge}} E_{\mathrm{Ge\,bulk}} - \Delta N_{\mathrm{Al}}\, E_{\mathrm{Al\,bulk}})/N_{\mathrm{Ge}}$$

and the formation energy per unit area, $\gamma$, which is calculated in the following way:

$$\gamma = (E_{\mathrm{Ge-Al}} - E_{\mathrm{backsideAl}} - N_{\mathrm{Ge}}\mu_{\mathrm{Ge}} - N_{\mathrm{Al}} E_{\mathrm{Al\,bulk}})/A$$

where $A$ is the area of the $(3 \times 3)$ mesh, equal to 0.64 $nm^2$, $E_{\mathrm{Ge-Al}}$ is the total energy of the considered model, $E_{\mathrm{backsideAl}}/A$ is the surface energy of the backside of the slab, $N_{\mathrm{Al}}$ and $N_{\mathrm{Ge}}$ the number of Al and Ge atoms in the slab, $\mu_{\mathrm{Ge}}$ the chemical potential of Ge and $E_{\mathrm{Al}}$ is the energy of a slab of same lateral size without Ge, whereas $\Delta N_{\mathrm{Al}}$ is the difference of number of Al atoms between the considered Ge/Al model and the bare Al slab. We have adopted an approach similar to the one used for comparing superstructures formed during Si growth on Ag(111):[46] we assume that the atomic mobility of the Al atoms is high enough to ensure that Al atoms in the surface reconstruction are at thermodynamic equilibrium with the bulk reservoir. This fixes the chemical potential for aluminum atoms to the bulk Al energy $E_{\mathrm{Al\,bulk}}$.

In Table 1 are shown the adsorption energies for Ge atoms in the different models. Whereas positive values are found for germanene models (first family) and for substitutional alloys with 9 atoms in the surface layer (third family), negative values are found for substitutional honeycomb layers with 8 atoms in the surface (second family). This indicates that Ge atoms can wet the Al surface by forming one of these latter structures, since their adsorption energy is lower than the bulk Ge energy.

| Surface layer | Interface layer | $E_{ad}$ (eV/at) | Figure |
|---|---|---|---|
| $Ge_8$ | $Al_9$ | 0.066 | S2m |
| $Ge_8$ | $Ge_1Al_8$ | 0.079 | S2n |
| $Ge_8$ | $Ge_2Al_7$ | 0.088 | S2o |
| $Ge_8$ | $Ge_3Al_6$ | 0.100 | S2p |
| $Ge_4Al_4$ | $Al_9$ | -0.043 | S2e |
| $Ge_4Al_4$ | $Ge_1Al_8$ | -0.031 | S2f |
| $Ge_4Al_4$ | $Ge_2Al_7$ | -0.027 | S2g |
| $Ge_4Al_4$ | $Ge_3Al_6$ | 0.011 | S2h |
| $Ge_5Al_3$ | $Al_9$ | -0.051 | S2i |
| $Ge_5Al_3$ | $Ge_1Al_8$ | -0.040 | S2j |
| $Ge_5Al_3$ | $Ge_2Al_7$ | -0.011 | S2k |
| $Ge_1Al_8$ | $Al_9$ | 0.094 | S2b |
| $Ge_3Al_6$ | $Al_9$ | 0.088 | S2c |
| $Ge_3Al_6$ | $Ge_1Al_8$ | 0.155 | S2d |
| $Ge_5Al_4$ | $Ge_1Al_8$ | 0.070 | S2l |

Table 1. Adsorption energy per Ge atom computed for different models from different families.

In Fig. 3, the formation energy $\gamma$ is displayed as a function of $\Delta\mu_{Ge}$, the deviation of $\mu_{Ge}$ with respect to $E_{Ge\,bulk}$. As a consequence, all curves appear as straight lines whose slope is given by the total number of Ge atoms $N_{\mathrm{Ge}}$ in the two planes. Mathematically, the models found to have negative $E_{\mathrm{ad}}$ values have a lower formation energy than the Al(111) surface energy for $\Delta\mu_{\mathrm{Ge}} = 0$.

Clearly the initial 1H model proposed for pure germanene on Al(111) ($Ge_8/Al_9$, red continuous line) is never favorable. Even if, for a certain range of germanium chemical potential, their surface energy is not far from the minimum one, the two best models found from analysis of the SXRD structure factors, $Ge_4Al_4/Ge_2Al_7$ (black dash-dot-dot line) and $Ge_8/Ge_2Al_7$ (red dash-dot-dot line), are also not the most stable ones. Indeed, for a large range of chemical potential, the most stable configurations found by DFT correspond to $Ge_5Al_3$ mixed honeycomb layer on a substitutional $Ge_nAl_{9-n}$ alloys. $Ge_5Al_3/Al_9$ is the most stable structure found for $-0.05\ \mathrm{eV} < \Delta\mu_{\mathrm{Ge}} < 0.02\ \mathrm{eV}$, $Ge_5Al_3/Ge_1Al_8$ is the most stable structure found for $0.02\ \mathrm{eV} < \Delta\mu_{\mathrm{Ge}} < 0.16\ \mathrm{eV}$, and $Ge_5Al_3/Ge_2Al_7$ is the most stable structure found for $0.16\ \mathrm{eV} < \Delta\mu_{\mathrm{Ge}} < 0.29\ \mathrm{eV}$. Above 0.29 eV, the most stable structure is a pure Ge layer on top of an alloyed plane, i.e., $Ge_8/Ge_3Al_6$. Thus, as $\Delta\mu_{\mathrm{Ge}}$ increases, DFT calculations show that the number of Ge atoms increases first in the plane below the surface, and then also in the honeycomb surface layer.

Substitutional alloys (blue, green and pink lines in Fig. 3) are never favored. In particular, it is interesting to note that the $Ge_1Al_8/Al_9$ configuration has always a high surface energy. This structure (blue line) corresponds to isolated Ge atoms, inserted in the surface plane. The excess energy of an inserted Ge atom, with respect to $\mu_{\mathrm{Gebulk}}$ is 0.094 eV (see Table 1). Thus, Ge atoms preferentially form ordered structures as soon as deposition starts, which corresponds to the experimental observations,[20] whereas, for Ge/Ag(111)[28] or Si/Ag(111),[47] isolated inserted atoms are observed at the beginning of growth.

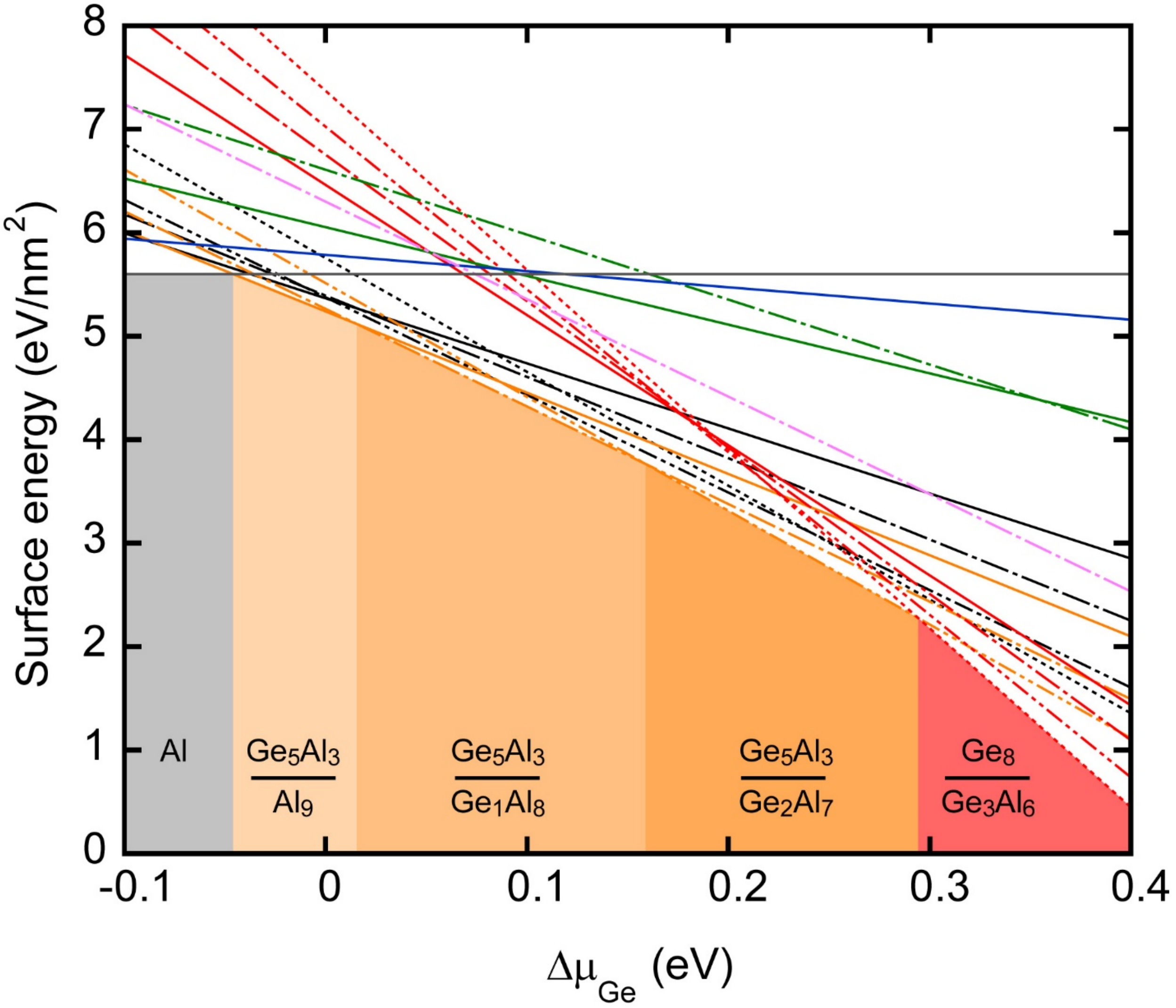

Fig. 3. Variation of the surface energy with respect to $\Delta\mu_{\mathrm{Ge}} = \mu_{\mathrm{Ge}} - E_{\mathrm{Ge\ bulk}}$. Gray continuous line: Al(111) surface. Other continuous lines: models on $Al_9$. Dash-dot lines: models on $GeAl_8$; Dash-dot-dot lines: models on $Ge_2Al_7$; dotted lines: models on $Ge_3Al_6$. Black: $Ge_4Al_4/Ge_nAl_{9-n}$; red: $Ge_8/Ge_nAl_{9-n}$; orange: $Ge_5Al_3/Ge_nAl_{9-n}$; blue: $Ge_1Al_8/Ge_nAl_{9-n}$; green: $Ge_3Al_6/Ge_nAl_{9-n}$; pink: $Ge_5Al_4/Ge_nAl_{9-n}$.

## Discussion

From the raw fitting of the SXRD data, we have obtained a family of models giving a good agreement with the experiments. From DFT, we have shown that a large number of these models have a better stability than a germanene layer on a pure Al(111) plane, and that the relative stability between the different models depends on the Ge chemical potential.

At this point, in order to finally determine the structure of the layer, we have computed the structure factors from the configurations relaxed by DFT. Only a scale factor and DW factors were used as free parameters. Corresponding $\chi^2$ for relevant configurations are given in Table 2. With $\chi^2 = 2.1$, the $Ge_4Al_4/Ge_2Al_7$ model clearly corresponds to the best configuration found. On the contrary, the $Ge_8/Ge_2Al_7$ model, with $\chi^2$ =10.2, must be rejected. Indeed, the atomic positions relaxed by DFT for $Ge_8/Ge_2Al_7$ are quite different from the ones found by initially fitting the structure factors, whereas in the case of $Ge_4Al_4/Ge_2Al_7$, they are very similar. All other models with 8 Ge atoms forming a honeycomb layer on a $Ge_nAl_{9-n}$ plane display poor agreement with the experiments. In particular, $Ge_8/Ge_3Al_6$, which was determined as the most stable structure for $\Delta\mu_{Ge} > 0.29$ eV, corresponds to $\chi^2 = 14.0$. Among the other models that display the lowest surface energy, the $Ge_5Al_3/Al_9$ and $Ge_5Al_3/Ge_1Al_8$ must also be rejected since they lead to $\chi^2 = 14.9$ and $\chi^2 = 12.4$ respectively. $Ge_5Al_3/Ge_2Al_7$ displays a better agreement with the experiments, with $\chi^2 = 6.1$.

As can be seen in table 2, most of the models relaxed by DFT correspond to $\chi^2 \geq 9.8$. For these models, some rods are correctly fitted, while other are badly reproduced. The four best models correspond to $\chi^2 \leq 6.1$. For these models, the variations of the intensity along all the rods are reasonably reproduced. These models are all very similar to $Ge_4Al_4/Ge_2Al_7$ and differ only by the modification of the chemical nature of one atom of the structure. Among them, $Ge_4Al_4/Ge_2Al_7$ gives the best fit to the structure factors, whereas $Ge_5Al_3/Ge_2Al_7$ is the most stable for $0.16$ eV $<$ $\Delta\mu_{Ge} < 0.29$ eV. The comparison between experimental and simulated structure factors is shown for these two models in Fig. 4.

| **Surface layer** | **Interface layer** | $\chi^2$ | **Figure** |
|---|---|---|---|
| $Ge_4Al_4$ | $Ge_2Al_7$ | 2.1 | S2g |
| $Ge_4Al_4$ | $Ge_3Al_6$ | 4.7 | S2h |
| $Ge_4Al_4$ | $Ge_1Al_8$ | 5.8 | S2f |
| $Ge_5Al_3$ | $Ge_2Al_7$ | 6.1 | S2k |
| $Ge_4Al_4$ | $Al_9$ | 9.8 | S2e |
| $Ge_8$ | $Ge_2Al_7$ | 10.1 | S2o |
| $Ge_5Al_3$ | $Ge_1Al_8$ | 12.4 | S2j |
| $Ge_8$ | $Ge_3Al_6$ | 14.0 | S2p |
| $Ge_5Al_3$ | $Al_9$ | 14.9 | S2i |
| $Ge_8$ | $Ge_1Al_8$ | 16.0 | S2n |
| $Ge_8$ | $Al_9$ | 26.6 | S2m |

Table 2. Comparison of the agreement between experiments and $Ge_mAl_{8-m}/Ge_nAl_{9-n}$ models relaxed by DFT.

For nearly all rods, the $Ge_4Al_4/Ge_2Al_7$ model reproduces perfectly the experimental structure factors. For some rods, such as $(02L)$ or $(20L)$, the agreement is less good, but experiments are not better fitted by the $Ge_5Al_3/Ge_2Al_7$ model for these specific rods. At this point, one may wonder if $\Delta\mu_{Ge}$ can be higher than 0.16 eV. For a system in equilibrium with the gas phase, the chemical potential is the same in the gas and in the adsorbed phases:

$$\mu_{Ge} = \mu_{0,gas} + kT\ln(P/P_0) = E_{ad} - Ts$$

where $P$ is the pressure of the gas, considered as ideal, and $s$ is the configuration entropy per atom of the adsorbed phase. For example, for $Ge_5Al_3/Ge_2Al_7$, $E_{ad} = E_{Ge\ bulk}$-0.011, whereas for nearly isolated Ge atoms, in the $Ge_1Al_8$/Al model, $E_{ad} = E_{Ge\ bulk}$+0.094. For a growing system, out of equilibrium, one expects an excess of chemical potential for the gas phase with respect to the adsorbed phase. In this case, however, one should keep the thermodynamical equilibrium between nearly isolated Ge atoms and the $(3 \times 3)$ growing regions. The chemical potential for the inserted atoms is $\mu_{Ge} = E_{Ge\ bulk} + 0.094 + kT\ln(c)$ where $c$ is the concentration of this diluted phase. Thus, it would be surprising that $\Delta\mu_{Ge}$ could be higher than 0.094 eV. From the thermodynamical diagram, only $Ge_5Al_3/Ge_1Al_8$ or $Ge_5Al_3/Al_9$ models should then correspond to the germanene layer. However, they display a poor agreement with the SXRD measurements ($\chi^2 = 12.4$ and $\chi^2 = 14.9$) respectively.

On the contrary, the difference of formation energy between the $Ge_4Al_4/Ge_2Al_7$ and the $Ge_5Al_3/Ge_1Al_8$ models is only 0.073 eV per unit cell. Moreover, as the formation energies are computed at 0K, whereas experiments are performed at 373 K, we cannot exclude that the difference of substrate temperature affects the relative stability of these phases. Furthermore, various models present adsorption energies that are quite similar (see Table 1). Since the thermal energy during growth is $kT = 0.032$ eV, we cannot completely exclude the fact that the $(3 \times 3)$ structure could result of a combination of the different models, in particular concerning the precise number of substitutional Ge atoms in the Al plane.

Finally, we want to point out that the formation of a two layers thick alloy explains why nucleation of $(3 \times 3)$ domains occurs near the step edges.[20] As extra Al atoms per unit cell are needed to build the structure, the flow of Al atoms is more easily obtained near a step edge than in the center of a terrace.

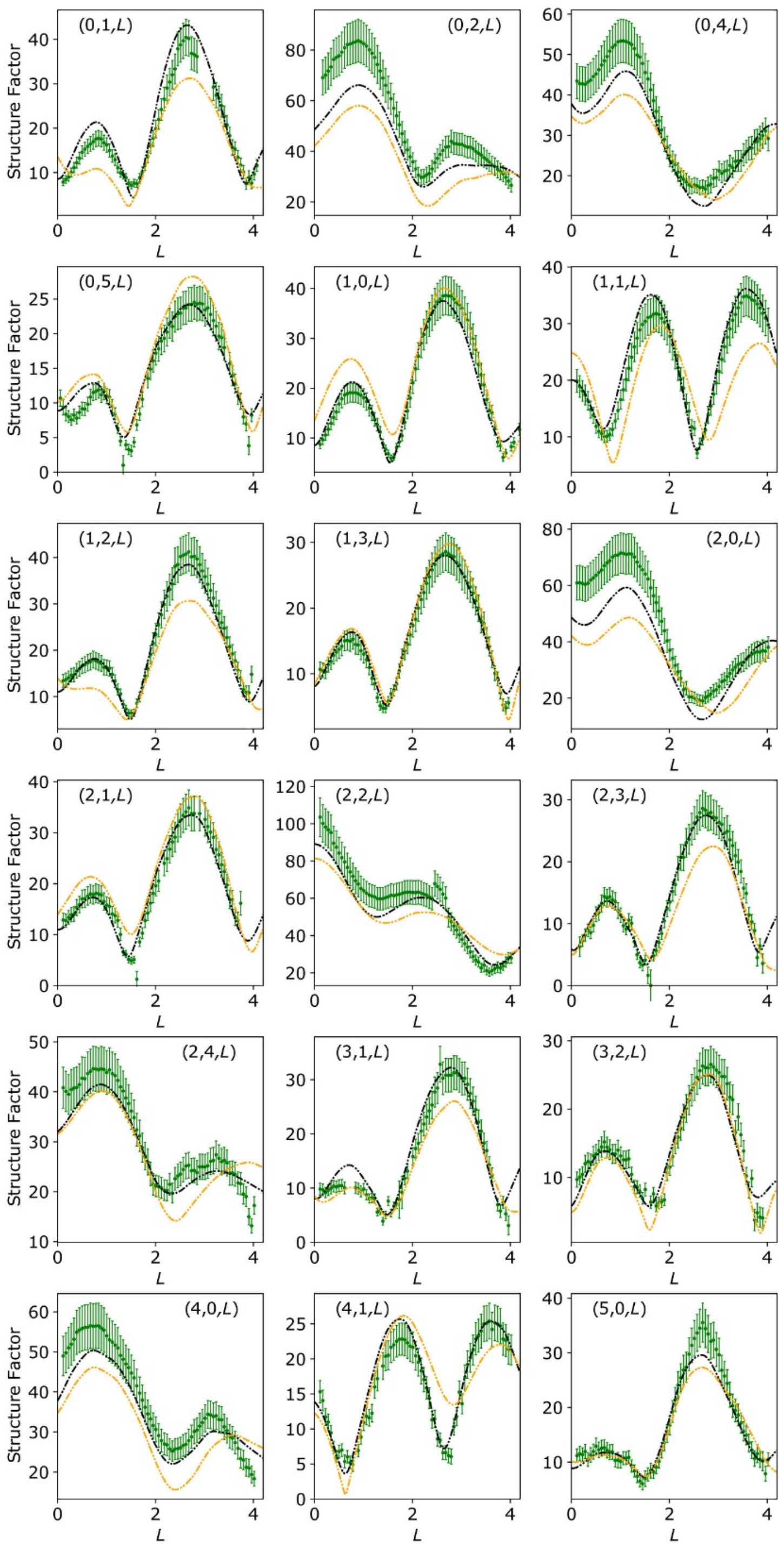

Fig. 4. Comparison between experimental (green dots) and simulated structure factors for the $Ge_4Al_4/Ge_2Al_7$ (black dash-dot-dot lines) and $Ge_5Al_3/Ge_2Al_7$ models (orange dash-dot-dot lines), along the different measured rods.

## Conclusion

In conclusion, our combined SXRD and DFT analysis of the germanene/Al(111) $(3 \times 3)$ reconstruction obtained by evaporating Ge on a substrate held at 373K demonstrates that this structure corresponds to a two-layers surface alloy. A pure germanene layer on a Al(111) substrate can be completely excluded. The best fit of the SXRD data corresponds to a $Ge_4Al_4/Ge_2Al_7$ structure where the surface plane has a honeycomb organization and the interface plane is a substitutional alloy. From a thermodynamical point of view, it has a formation energy very close to the one of $Ge_5Al_3/GeAl_8$. This latter one is the most stable structure found by DFT at 0K for $0.02\ \text{eV} < \Delta\mu_{Ge} < 0.16$ eV, which also corresponds to the expected experimental condition, i.e., a small excess of chemical potential with respect to the bulk energy of Ge.

ASSOCIATED CONTENT

**Supporting Information**.

The following files are available free of charge.

Details of the minimization procedure used for exploring all configurations, relaxed configurations of the models computed by DFT (PDF).

AUTHOR INFORMATION

**Corresponding Author**

Geoffroy Prévot – Sorbonne Université, Centre National de la Recherche Scientifique, Institut des NanoSciences de Paris, INSP, F-75005 Paris, France; https//orcid.org/0000-0001-7960-5587; Email: prevot@insp.jussieu.fr

**Authors**

Kai Zhang – Sorbonne Université, Centre National de la Recherche Scientifique, Institut des NanoSciences de Paris, INSP, F-75005 Paris, France

Davide Sciacca – Univ. Lille, CNRS, Centrale Lille, Univ. Polytechnique Hauts-de-France, Junia-ISEN, UMR 8520 - IEMN, F-59000 Lille, France

Marie-Christine Hanf – Université de Haute Alsace, CNRS, IS2M UMR7361, F-68100 Mulhouse, France;  Université de Strasbourg, France

Romain Bernard – Sorbonne Université, Centre National de la Recherche Scientifique, Institut des NanoSciences de Paris, INSP, F-75005 Paris, France; https//orcid.org/0000-0002-2944-142X

Yves Borensztein – Sorbonne Université, Centre National de la Recherche Scientifique, Institut des NanoSciences de Paris, INSP, F-75005 Paris, France; https://orcid.org/0000-0002-1570-732X

Andrea Resta – Synchrotron SOLEIL, L'Orme des Merisiers Saint-Aubin, BP 48 91192 Gif-sur-Yvette Cedex, France

Yves Garreau – Synchrotron SOLEIL, L'Orme des Merisiers Saint-Aubin, BP 48 91192 Gif-sur-Yvette Cedex, France; Université de Paris, Laboratoire Matériaux et Phénomènes Quantiques, CNRS, F-75013, Paris, France

Alina Vlad – Synchrotron SOLEIL, L'Orme des Merisiers Saint-Aubin, BP 48 91192 Gif-sur-Yvette Cedex, France

Alessandro Coati – Synchrotron SOLEIL, L'Orme des Merisiers Saint-Aubin, BP 48 91192 Gif-sur-Yvette Cedex, France; https://orcid.org/0000-0001-7732-5688

Isabelle Lefebvre – Univ. Lille, CNRS, Centrale Lille, Univ. Polytechnique Hauts-de-France, Junia-ISEN, UMR 8520 - IEMN, F-59000 Lille, France

Mickael Derivaz – Université de Haute Alsace, CNRS, IS2M UMR7361, F-68100 Mulhouse, France;  Université de Strasbourg, France; https://orcid.org/0000-0002-7802-6492

Carmelo Pirri – Université de Haute Alsace, CNRS, IS2M UMR7361, F-68100 Mulhouse, France; Université de Strasbourg, France; https://orcid.org/0000-0002-3629-1044

Philippe Sonnet – Université de Haute Alsace, CNRS, IS2M UMR7361, F-68100 Mulhouse, France; Université de Strasbourg, France; https//orcid.org/

Régis Stephan – Université de Haute Alsace, CNRS, IS2M UMR7361, F-68100 Mulhouse, France; Université de Strasbourg, France; https//orcid.org/

ACKNOWLEDGMENTS

This study is financially supported the French National Research Agency (Germanene project ANR-17-CE09-0021-03). K.Z. is supported by the Chinese Scholarship Council (CSC contract 201808070070). The authors would like to acknowledge the High Performance Computing Center of the University of Strasbourg for supporting this work by providing scientific support and access to computing resources. Part of the computing resources were funded by the Equipex Equip@Meso project (Programme Investissements d'Avenir) and the CPER Alsacalcul/Big Data.

TOC Graphic

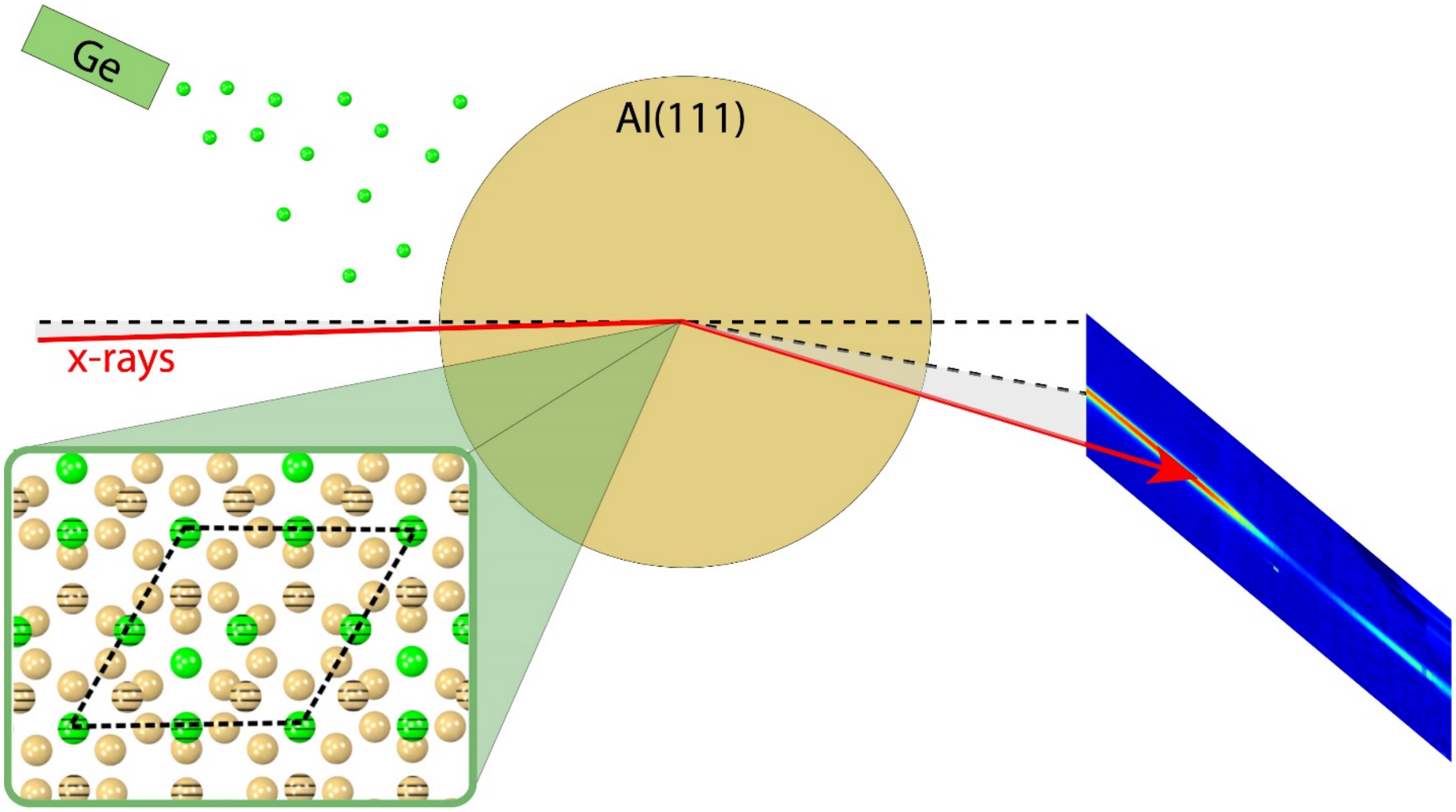